\pdfoutput=1
\documentclass[pre,aps,twocolumn,showpacs,floatfix,nofootinbib,superscriptaddress]{revtex4-1}

\usepackage{url}
\usepackage[T1]{fontenc}
\usepackage[utf8]{inputenc}

\usepackage{subfigure}
\usepackage{amssymb}
\usepackage{amsfonts}
\usepackage{amsmath}
\usepackage{amsthm}
\usepackage{epsfig}
\usepackage{graphicx}
\usepackage{cancel}

\usepackage{float}
\usepackage{placeins}
\usepackage{threeparttable}

\usepackage{xcolor}
\usepackage{color}
\usepackage[hidelinks,unicode=true]{hyperref}
\usepackage{comment}
\usepackage[normalem]{ulem}
\hypersetup{colorlinks=true,
 	linkcolor=blue,
 	urlcolor=blue,
 	citecolor=blue,
 	pdfhighlight=/N
 }
\renewcommand{\vec}{\mathbf}

\newcommand{\tr}{ }
\begin{document}
\title{ Scaling, Fractal Dynamics and Critical Exponents: Application in a non-integer dimensional ising model}
\author{Henrique A. de Lima}
\email{henrique\_adl@hotmail.com}
\address{International Center of Physics, Institute of Physics, University of Brasilia, 70910-900, Brasilia, Federal District, Brazil}
\author{Ismael S. S. Carrasco}
\email{ismael.carrasco@unb.br}
\address{International Center of Physics, Institute of Physics, University of Brasilia, 70910-900, Brasilia, Federal District, Brazil}
\author{Marcio Santos}
\email{marcio.santos@ufsc.br }
\address{International Center of Physics, Institute of Physics, University of Brasilia, 70910-900, Brasilia, Federal District, Brazil}
\address{Instituto de F\'{\i}sica, Universidade Federal de Santa Catarina, Florian\'{o}polis, SC, Brazil}
\author{Fernando A. Oliveira}
\email{faooliveira@gmail.com}
\address{International Center of Physics, Institute of Physics, University of Brasilia, 70910-900, Brasilia, Federal District, Brazil}
\begin{abstract}
  Moving beyond simple associations, researchers need tools to quantify how variables influence each other in space and time. Correlation functions provide a mathematical framework for characterizing these essential dependencies, revealing insights into causality, structure, and hidden patterns within complex systems. In physical systems with many degrees of freedom, such as gases, liquids, and solids, a statistical analysis of these correlations is essential. For a field $\Psi(\vec{x},t)$ that depends on spatial position $\vec{x}$ and time $t$, it is often necessary to understand the correlation with itself at another position and time $\Psi(\vec{x}_0,t_0)$. This specific function is called the autocorrelation function. In this context, the autocorrelation function for order--parameter fluctuations, introduced by Fisher~\cite{Fisher64} ( M. E. Fisher, Journal of Mathematical Physics 5, 944322 (1964)), provides an important mathematical framework for understanding the second-order phase transition at equilibrium. However, his analysis is restricted to a Euclidean space of dimension $d$, and an exponent $\eta$ is introduced to correct the spatial behavior of the correlation function at $T=T_c$. In recent work, Lima et al~\cite{Lima24} (Lima et al, Phys. Rev. E 110, L062107 (2024)) demonstrated that at $T_c$ a fractal analysis is necessary for a complete description of the correlation function. In this study, we investigate the fundamental physics and mathematics underlying phase transitions, emphasizing the deep interplay between scaling behavior, critical exponents, and fractal geometry. In particular, we show that the application of modern fractional differentials allows us to write down an equation for the correlation function that recovers the correct exponents below the upper critical dimension. We obtain the exact expression for the Fisher exponent $\eta$. Furthermore, we examine the Rushbrooke scaling relation, which has been questioned in certain magnetic systems, and, drawing on results from the Ising model, we confirm that both our relations and the Rushbrooke scaling law hold even when $d$ is not an integer.

\end{abstract}
\maketitle
\section{Introduction}
\label{Int}
Phase transitions represent abrupt or gradual transformations in the properties of physical systems due to changes in external parameters, such as temperature or pressure \cite{Kardar07}. In general, this phenomenon involves structural reorganization at the particle level. They are classified as first-or second-order, depending on the continuity of thermodynamic variables or response functions, respectively.

On the other hand, since Mandelbrot's pioneering work~\cite{Mandelbrot82}, it has become evident that fractals play a fundamental role in physics, particularly in condensed matter and statistical physics because, near phase transitions, concepts such as renormalization and scaling are strongly connected~\cite{Cardy96}. In fact, the scaling behavior observed at critical points is a manifestation of fractal geometries \cite{Suzuki8, Kroger00,Devakul19}, highlighting the deep connections between scaling, renormalization, and fractals. In this work,  we focus on the second-order phase transition at equilibrium and further strengthen these connections, clarifying concepts such as the applicability and limitations of mean-field theory. 

A key quantity that deserves in-depth analysis is the autocorrelation function
\begin{equation}
\label{G0}
    G(r)=\langle \psi(\vec{r}+\vec{x})\psi(\vec{x})\rangle,
\end{equation}
where $\psi(\vec{x})$ represents the fluctuation of the order parameter at the position $\vec{x}$, and $\langle \cdots\rangle$ denotes the average taken at equilibrium over the lattice in a real space with dimension $d$. Ornstein and Zernike's seminal work~\cite{Ornstein14} on the density fluctuations in fluids led us to the conclusion that, in the continuous limit, only long-wavelength fluctuations contribute to the system's collective behavior. This implies that (\ref{G0}) can be described by an equation of the form ~\cite{Kardar07} 
\begin{equation}
\label{G}
 (-\nabla^2 +\rho^{-2})G(r)=\delta^{(d)}(r)   
\end{equation} 
where $\rho$ is the correlation length. Note that this equation applies to an infinite medium with no defects. Therefore, we will consider spherical symmetry in the solution below. The correlation length diverges as the temperature $T$ approaches the critical temperature $T_c$, following a well-known scaling behavior~\cite{Kardar07,Cardy96}
\begin{equation}
\label{rhodivergence}
\rho \propto |T-T_c|^{-\nu}.
\end{equation}
The solution of (\ref{G}) is given by $G(r)\propto r^{2-d} \exp(-r/\rho)$. This result implies that, in the limit $\rho \rightarrow \infty $, corresponding to the critical point, the correlation function scales as $G(r) \propto r^{2-d} $.

However, this mean-field solution fails at the critical point. For $d=2$, it predicts a constant correlation, independent of the distance $r$ between the sites, while for $d<2$, which we shall discuss here, the result is even worse, as it predicts a correlation that increases with distance. Therefore, the solution for $d \leq 2$ is obviously unphysical. Moreover, data for X-ray scattering experiments on simple fluids exhibit deviation from this predicted exponent\cite{Fisher64}. To account for this discrepancy, a correction was applied~\cite{Fisher64}, introducing the Fisher exponent $\eta$
\begin{equation}
\label{G2}
G(r) \propto
\begin{cases}
r^{2-d} \exp(-r/\rho) , &\text{ if~~ } r>\rho,\\
r^{2-d-\eta}, &\text{ if~~ }  r \ll\rho.\\
\end{cases}
\end{equation}
Along with this correction, a new scaling relation was proposed\cite{Fisher64}
\begin{equation}
\label{gamma}
\gamma =(2-\eta)\nu,
\end{equation}
where $\gamma$ is the susceptibility exponent. 
 
The above relation may be verified for the $2d$ Ising  model, which is equivalent to the plane square lattice gas, where exact values $\gamma=7/4$ and $\nu=1$ are established.  Kaufman and Onsager~\cite{Kaufman49} further derived the correlation function $G(r) \propto r^{-1/4} $, yielding $\eta=1/4$. These exponents agree with Eq. (\ref{gamma}) and have been rigorously confirmed across diverse systems over the past seven decades. While the necessary inclusion of $\eta$ in Eq. (\ref{G2}) is unequivocal, and the main reason behind this modification was not conceptually clear.

\section{ A Fractal Analysis }

Trying to describe the origins of this correction, in a recent work, Lima et al~\cite{Lima24} conducted a thorough analysis of the correlation function (\ref{G}). The authors drew inspiration from three key observations: 
\begin{enumerate}
\item Fluctuation-Dissipation Theorem (FDT): The correlation functions represent a simple form of the FDT (see the article collection~\cite{Nowak22} and the  reviews \cite{Oliveira19,GomesFilho25}), which are directly related to susceptibility \cite{Goldenfeld18}. The FDT is known to break down when ergodicity is violated \cite{Costa03,Wen23}, a phenomenon widely studied in structural glasses \cite{Grigera99,Ricci-Tersenghi00,Crisanti03,Barrat98,Bellon02,Bellon06}, random-exchange Heisenberg chains \cite{Vainstein05}, proteins \cite{Hayashi07}, KPZ dynamics \cite{Kardar86,Rodriguez19}, mesoscopic radioactive heat transfer \cite{Perez-Madrid09,Averin10}, and ballistic diffusion \cite{Costa03,Costa06,Lapas07,Lapas08};
\item Response Functions in Growth Phenomena: Significant progress has been made in understanding response functions in growth phenomena by assuming an underlying fractal dynamics \cite{Barabasi95,Feder22, GomesFilho21b,Anjos21,Luis22,Luis23,GomesFilho24};
\item  Scaling and Fractal Geometries:  The relation between the order parameter exponent $\beta$ and
the fractal dimension of the ordered phase $d_f$ was first proposed by Suzuki \cite{Suzuki8}
\begin{equation}
\label{dl}
d_f=d-\frac{\beta}{\nu}\;.
\end{equation}

For the critical point of the percolation model, this fractal structure corresponds to the infinite percolating cluster\cite{Grimmett06,Cruz23}, while for general systems it is associated with the largest ordered cluster~\cite{Kroger00}. 
\end{enumerate}

\begin{figure}
    \centering
    \includegraphics[width=0.97\linewidth]{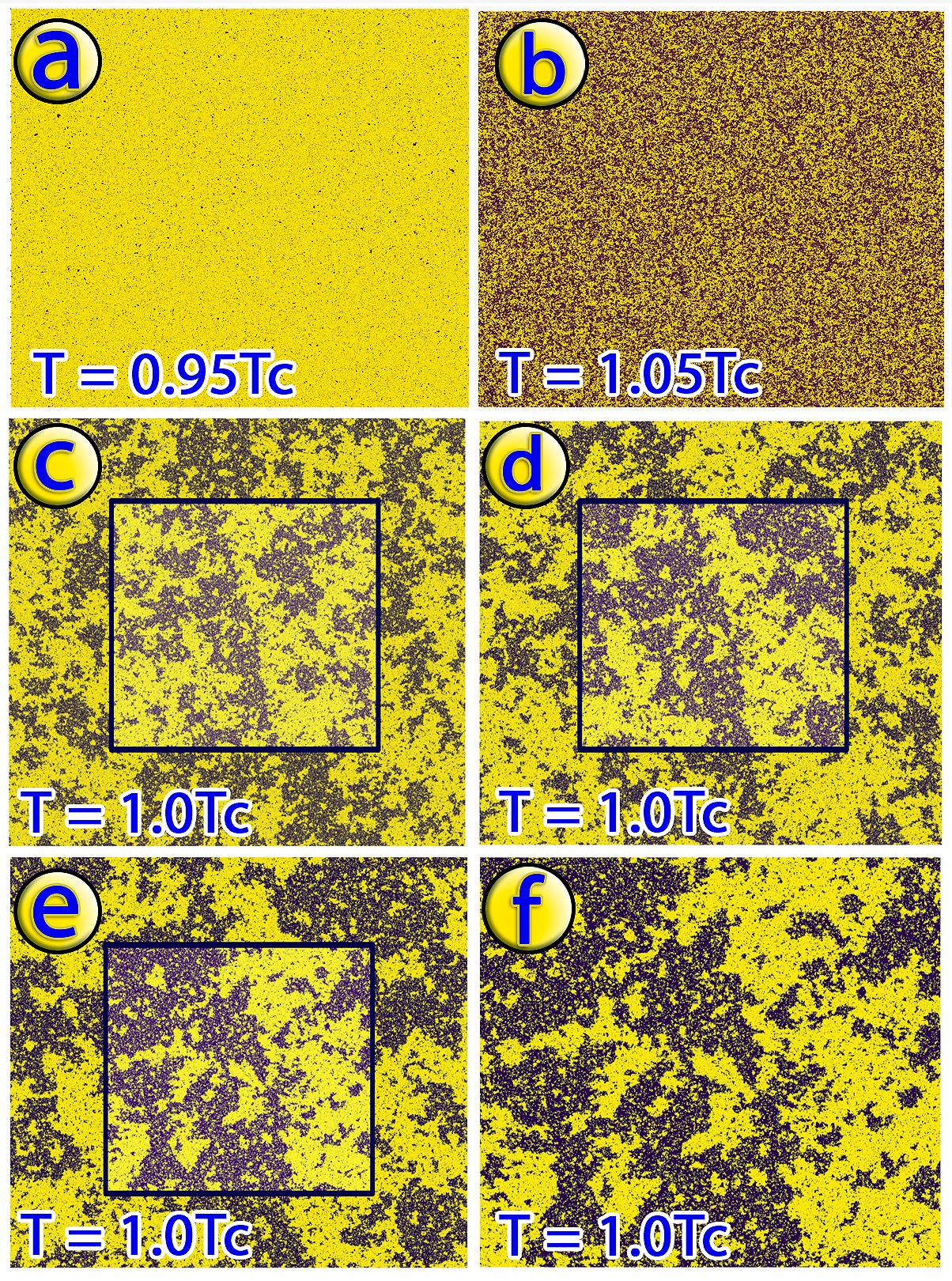}
    \caption{Snapshots of spin configurations in the $2d$ Ising model at various temperatures, generated via Metropolis Monte Carlo simulations. Yellow and dark blue represent spins $+1$ and $-1$, respectively. Panel a) shows predominantly ferromagnetic order with minor disorder at temperature $T = 0.95 T_{c}$. In b) the system is in a fully disordered paramagnetic phase with magnetization $m = 0$. c)-f)  $T = T_c$ ($m=0$), where the correlation length $\rho \to \infty$. (c) Critical state overview; (d)-(f) sequential 2$\times$-magnified snapshots of the region inside the square in the previous panel. At criticality, the system exhibits fractal (scale-invariant) spin clusters and divergent correlation length parameter $\rho(T)$.}
    \label{grids_}
\end{figure}

Based on these concepts, the authors analyzed the correlation function within the following basic ideas:
\begin{enumerate}

\item  Mean field theory proves inefficient for analyzing the order parameter at the critical point because it suppresses fluctuations when they are relevant. Notably, according to Eq. (\ref{rhodivergence}), $\rho$ diverges as $T \rightarrow T_c$, indicating long range correlations. This is associated to the emergence of a scale-invariant fractal structure where fluctuations manifest across all length scales. Consequently, this function should provide a precise description if the appropriate geometry is used.

 \item In physics, fractal geometry can emerge from predefined rules, such as anomalous infiltration in partially saturated porous media~\cite{Voller25} or fractal circuits~\cite{Chen17,Boyle07,Anjos24}.  However, when derivatives are defined in a Euclidean space, a fractional derivative may be necessary to extend the analysis to a fractal medium, as in the fractional heat equation~\cite{Angulo00} or the fractional Fokker-Planck equation~\cite{Metzler99,Barkai01,Sokolov01}. This is the case for Eq. (\ref{G}), as suggested by Fig. \ref{grids_}, where we show some representations of a $2d$ Ising lattice. At critical temperature $T_{c}$, the system displays a fractal structure (scale--invariant) of spin clusters, along with the divergence of $\rho(T)$. The variation of $G(r)$, and thus of $\nabla^2 G$, is more relevant at the fractal edge of the clusters, which is where Eq.(\ref{G}) is nontrivial.

\item As $T \rightarrow T_c$ and $\rho$ diverge according Eq. (\ref{rhodivergence}), the correlation extends over a long range thanks to the emergence of a fractal structure, which may be better explained by the use of a fractal Laplacian.  Then, we replace Eq.\ (\ref{G}) by 
\begin{equation}
\label{G3}
(-\nabla^2)^\zeta G(r)=\delta^{(d_R)}(r),
\end{equation}
where $d_R$ is a fractal dimension associated with the Riesz fractional derivative of order $\zeta$. The preference for the Riesz fractional derivative is \tr{due} the existence of a well\tr{--}defined Laplacian for it and the identity ~\cite{Muslih10,Muslih10b}
\begin{equation}
\label{SG3}
(-\nabla^2)^\zeta \left(\frac{1}{|\vec{r}-\vec{r'}|^{d_R-2\zeta}} \right) = C_{\zeta,d_R}  \delta^{(d_R)}(\vec{r}-\vec{r'}),
\end{equation}
where  $\delta^{(d_R)}(\vec{r}-\vec{r'})$ is a $\delta$-function in a space of fractal dimension $d_R$ and $C_{\zeta,d_R}=[4^{\zeta} \pi^{d_R/2} \Gamma(\zeta)]/[ \Gamma(d_R/2-\zeta)]$ is a constant. Another important mathematical aspect is the nonlocality of the fractional derivative, which is in agreement with the  collective phenomenon that is a phase transition.

Comparing Eq. (\ref{G2}), in the limit $\rho \rightarrow \infty$, with (\ref{G3}) and  (\ref{SG3}) we get $ \eta=2(1-\zeta)+d_R-d$. Finally using a Gauss theorem for fractal distributions, the Fisher exponent is derived exactly as~\cite{Lima24}
 
\begin{equation}
\label{eta}
 \eta=d-d_R=1-\zeta.
\end{equation}

\end{enumerate}
Consequently, the Fisher exponent in the correlation function $G(r)$ quantifies the deviation of the correlation fractal dimension $d_R$ (the  Riesz fractional dimension), from the Euclidean integer dimension $d$. As demanded by the Riesz fractional derivative~\cite{Muslih10,Muslih10b}  $d-1 \leq d_R\leq d$. This is consistent with the fact that $\eta \geq 0$.

Furthermore, following the approach in \cite{Lima24}, combining Eqs. (\ref{gamma}), (\ref{dl}), (\ref{eta}), (\ref{alpha}), and (\ref{rush}) yields the relationship
\begin{equation}
\label{dl2}
    d_R=2(d_f-1).
\end{equation}

At this point, we emphasize that the fractional equation (\ref{G3}) was the crucial missing element that links together the exponent relations, allowing the Fisher exponent to be expressed as a function of the Riesz fractal dimension, Eq. (\ref{eta}). The exponents obtained from it are in agreement with exact and approximated results in the literature, see the following section. It is also possible to measure $\beta$ and $\nu$ and from this obtain $d_f$ and $d_R$ or even to obtain the fractal dimension experimentally\footnote[1]{Although Eq. (\ref{eta}) is very recent, there are some methods of measuring fractal dimensions~\cite{Lin89,Nayak17,Dumouchel05,Pham23}, some of which can be used to measure $d_R$.}.

Given the importance of such exact relationships, our goal for this manuscript is as follows.
\begin{enumerate}
    \item Generalize these  results to non-integer dimensions $d$;
    \item  Verify these results in extreme cases where scaling relationships break down.
\end{enumerate}


\section{Non-integer dimensional lattice }

Retracing the derivation of Eqs. (\ref{eta}) and (\ref{dl2}), we conclude that these relationships remain valid for non-integer dimensions $d$. However, fractional dynamics is subtle, particularly those related to fractional differential geometry. {Thus, to check if those relationships really work for non-integer dimensions,} we will analyze existing data in the literature. We will start with the Ising model in a lattice with non-integer dimension in the range $1 \leq d \leq 4$. 

\begin{table*}[t]

\begin{tabular}{c|cccccc} \hline\hline
$d$     & $\beta$                           & $\nu$                             & $\eta^*$                            & $d_f$ & $d_R$ & $\eta$ \\ \hline
4     & 1/2                            & 1/2                            & 0                              &  3   &  4 & 0\\
3.75  & $0.458355(2)$ & $0.523405 (15)$ & $0.001435(1)$ &  $2.87428(2)$  &  $3.74856(4)$ &  $0.001435(4)$  \\
3.5   & $0.41600(2)$  & $0.55215(2)$  & $0.00685(2)$  &  $2.74658(1)$  &  $3.49316(2)$ &  $0.00683(2)$ \\
3.25  & $0.3723(1)$   & $0.5873(1)$   & $0.0180(1)$   &  $2.61608(6)$  &  $3.2321(1)$ & $0.0178(1)$ \\
3     & $0.3270(15)$   & $0.6310(15)$   & $0.0375(25)$   &  $2.481(1)$  & $2.963(3)$ & $0.036(3)$ \\
2.75  & $0.2800(30)$   & $0.686(3)$     & $0.067(6)$     &  $2.341(3)$  &  $2.683(5)$ & $0.066(5)$ \\
2.5   & $0.2305(3)$   & $0.758(5)$     & $0.11(1)$      &  $2.195(2)$  & $2.391(3)$ & $0.108(3)$ \\
2.25  & $0.1790(25)$   & $0.857(6)$     & $0.17(1)$      &  $2.041(2)$  & $2.082(4)$ & $0.167(4)$ \\
2     & 1/8     & 1      & 1/4   &  $15/8$  & $7/4$ &  1/4 \\
1.875 & $0.097(3)$     & $1.10(1)$      & $0.30(3)$      &  $1.786(2)$   & $1.573(4)$ & $0.301(4)$  \\
1.75  & $0.068(6)$     & $1.23(3)$      & $0.35(5)$      &  $1.695(3)$  & $1.389(7)$ & $0.361(7)$ \\
1.65  & $0.045(10)$    & $1.37(7)$      & $0.40(10)$     &  $1.617(5)$  & $1.23(1)$ & $0.41(1)$ \\
1.5   & $0.01(15)$    & $1.67(20)$     & $0.50(15)$  &  $1.494(8)$  & $0.99(2)$ &  $0.51(2)$ \\
1.375 & $-0.02(3)$     & $2.1(5)$       & $0.55(25)$     &  $1.38(1)$  & $0.79(3)$ & $0.61(3)$ \\
1.25  & $-0.05(5)$     & $3.0(15)$      & $0.65(35)$     &  $1.26(2)$   &  $0.53(5)$ & $0.71(5)$ \\
1     & 0                              & $\infty$       & 1                              &  1  & 0 & 1 \\ \hline\hline
\end{tabular}
\caption{Values of critical exponents for  the $d$ dimensional Ising model universality class. $\beta$, $\nu$ and $\eta^*$, from ref  \cite{Guillou87}, we use $\eta*$ to distinguish from $\eta$ obtained here from Eq.\ (\ref{eta}), $d_f$ from Eq. (\ref{dl}), $d_R$ from  Eq. (\ref{dl2}).}
\label{Table1}

\end{table*}


Before proceeding with our numerical analysis, we draw attention to the fractional exponents obtained in percolation clusters, as described, for instance, by Suzuki~\cite{Suzuki8} and Coniglio\cite{Coniglio89}. The exponent $d_R$ introduced here represents a new fractal dimension, distinct from those previously reported~\cite{Suzuki8,Coniglio89,Kroger00}.  However, Eq. (\ref{dl2}) established a connection between them.

In Table \ref{Table1}, we present the values of $d_f$ and $d_R$ for the $d$ dimensional Ising model, along with the corresponding critical exponents. We take the exponents $\beta $, $\nu$, and $\eta*$ from the literature. Here we use $\eta*$ to distinguish from $\eta$ obtained from (\ref{eta}), while $d_f$ is calculated from Eq. (\ref{dl}) and $d_R$  from (\ref{dl2}). For $d = 4$, we consider mean-field exponents, and for $d = 3$, we use the results of Pelissetto and Vicari~\cite{Pelissetto02}. For $d=2$, we use the exact critical exponents of the Ising model. In non--integer dimensions ($1<d<4$), we use  values obtained by Guillou and Justin~\cite{Guillou87} through their resummation method applied to the Wilson-Fisher $\epsilon$  expansion (where $\epsilon=4-d$), which is based on Borel transformation and conformal mapping. Their results  match the exact values of the Ising model at $d=2$ and are consistent with approximated values for $d=3$. For the more imprecise region ($1<d<2$), Novotny~\cite{Novotny92} independently confirmed these estimates. Finally, for $d=1$, we use the critical exponents from the near--planar interface model~\cite{Wallace79} and the droplet model~\cite{Bruce981,Bruce83}.

The computed values of $d_f$ agree with the exact results of Coniglio~\cite{Coniglio89} for $d=2$, and $4$. The values of $\eta$ obtained using our method begin to deviate from $\eta*$ as $d$ decreases, see reference ~\cite{Guillou87}. This is expected, given that Guillou and Justin~\cite{Guillou87} reported lower precision in their estimates for small values of $d$.

The previous results support the extension of Eqs. (\ref{eta}) and (\ref{dl2}) to non-integer dimensions. Furthermore, a detailed analysis of the data in the range $1<d<2$ reveals that precision decreases together with dimension, particularly for the value of $\eta^*$. However, the prediction from Eq. (\ref{eta}) leads to a much higher precision, highlighting the utility of this equation. 

 {
\begin{figure}[htbp] 
	\centering
	\includegraphics[width=0.7\columnwidth]{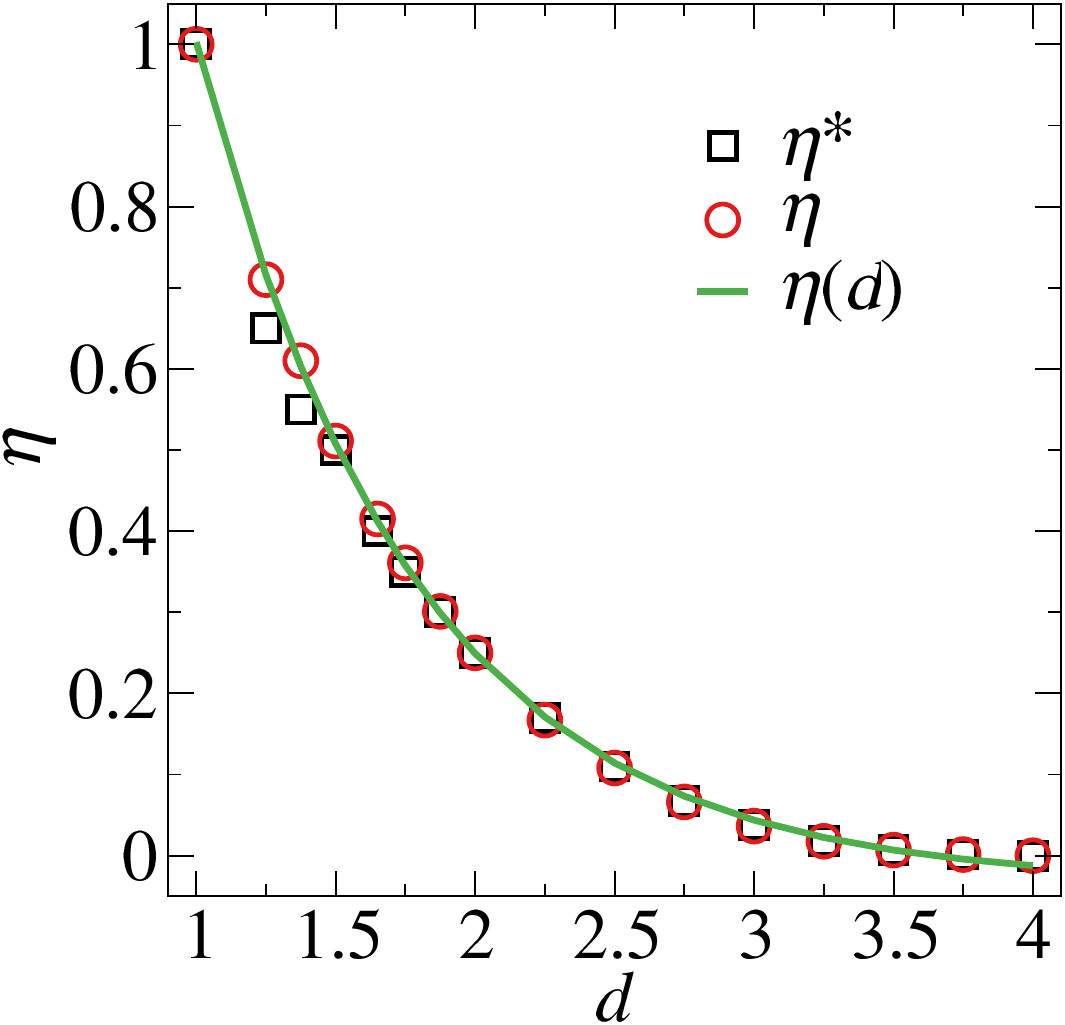}
        \caption{  { $\eta$ as function of $d$. Comparison between the values of $\eta^*$, here  $\eta^*$ is  $\eta$ from the literature} and the values $\eta$ calculated using Eq. \ref{eta} for dimensions in the range $1<d<4$. All data points are taken from Table \ref{Table1}, while the green curve represents the proposed function $\eta(d)$ given by Eq. (\ref{curve}).} 
	\label{figfractal}
\end{figure}
}

From  Table \ref{Table1} we adjust the function
\begin{equation}
\label{curve}
    \eta(d)=\frac{4-d}{5d-2} \left[1-\frac{1772(d-1)(d-2)}{625(5d-2)} \right].
\end{equation}
where we use the values of $\eta$ for integer dimensions,  being it  approximate~\cite{Pelissetto02} for $d=3$ and  exact for  $d=1,2,4$. These values are common to both $\eta$ and $\eta^*$

In Fig. \ref{figfractal}, we present $\eta$ as a function of $d$ from Table 1 to facilitate the comparison between the values of $\eta^*$ from the literature and the $\eta$ calculated using Eq. (\ref{eta}). The graph highlights the overall agreement between these values Additionally, the empirical function given by Eq. (\ref{curve}) is shown, which agrees well with the data points for $d \geq 2$, particularly with the values predicted by Eq. (\ref{eta}). For $1<d<2$ the computed values $\eta$ and the function $\eta(d)$,  Eq. (\ref{curve}), agree with each other, while they deviate slightly from $\eta*$.  This suggests that the proposed function serves as a reliable empirical rule for predicting the value of the exponent $\eta$ for dimensions in the range $1 \leq d \leq 4$.

\begin{table}[]
\begin{tabular}{c|cccc}
\hline\hline
$d$     & $\alpha$       & $\gamma$    & $\Sigma^*$  & $\Sigma$ \\ \hline
2     & 0          & 1.75     & 2    &  2  \\
1.875 & -0.0625(1) & 1.862(1) & 1.993(7) & 2.00(3)\\
1.75  & -0.15(5)  & 1.99(4)  & 1.97(5) & 2.00(2) \\
1.65  & -0.3(1)      & 2.11(8)  & 1.93(6) & 2.00(3) \\
1.5   & -0.5(3)  & 2.35(2)  & 1.89(4) & 2.0(3) \\
1.375 & -0.9(7)    & 2.6(4)   & 1.67(25) & 2.0(1)    \\
1.25  & -1.75(1.85)   & 3.0(1)   & 1.15(1) & 2.0(3) \\
1     & $-\infty$   & $\infty$ & $?$ & 2\\ \hline\hline

\end{tabular}

\caption{Values of critical exponents for  the $d$ dimensional Ising model universality class.  $\beta$, $\gamma$ and $\nu$ from~\cite{f.y.wu} to obtain $\alpha$, from Eq. (\ref{alpha}), 
$\Sigma^*= \alpha +2\beta+\gamma$ and  $\Sigma= \alpha +2\beta+\gamma$ from our values.}
\label{Table2}
\end{table}

{\bf Scaling - } Scaling is one of the simplest and most general mechanisms in physics. In the context of phase transitions, it enables us to obtain relationships between critical exponents, providing $N-2$ scaling relations for each set of $N$ exponents. For instance, the critical exponent of the correlation length $\nu$ is related to the critical exponent of specific heat $\alpha$ through the hyperscale relation~\cite{Cardy96}
\begin{equation}
\label{alpha}
\alpha=2-d\nu,
\end{equation}
thus associating a thermodynamic variable with the divergence of the correlation length.  We have as well the Fisher scaling relation Eq (\ref{gamma})
 and the Rushbrooke equality~\cite{Cardy96}
\begin{equation}
\label{rush}
\alpha +2\beta +\gamma=2.
\end{equation}
In the previous work \cite{Lima24}, to obtain the relation (\ref{dl2}), some of us have used Eq. (\ref{dl}) together with the scaling relations above. While the Rushbrooke equality (\ref{rush}) has been widely validated in equilibrium phase transitions, it was recently questioned by Fytas {\bf et al} for the random-field Ising model~\cite{Fytas16}. If this equality fails, Eq. (\ref{dl2}) would also break down, however,  Eq. (\ref{eta}) would remain valid. {Given that} {deviations may} occur in the region $1<d<2$, we analyze these values in Table \ref{Table2}.

In Table \ref{Table2}, we use the values of $\beta$, $\gamma$ and $\nu$ from~\cite{f.y.wu} to calculate $\alpha$ using the equation. (\ref{alpha}) and $\Sigma^*= \alpha +2\beta+\gamma$. Alternatively, $\Sigma$ is computed using the same expression, but with our results from Table \ref{Table1} and the relation $\alpha+\gamma=2+(2-d-\eta)\nu$, derived from the sum of Eqs. (\ref{alpha}) and (\ref{gamma}).  The table clearly shows that as the dimension decreases from $d=2$ to $1$, $\Sigma^*$ deviates from $2$, while $\Sigma$ remains constant within the error bounds.  This distinction is crucial for $d=1$, where $2-d-\eta=0$.   Thus we take $(2-d-\eta)\nu=0$, before  the  limit $\nu \rightarrow \infty$ and we obtain the correct result $\Sigma=2$, resolving the indeterminacy  in $\Sigma^*$.

Moreover, incorporating $\beta$ from (\ref{dl}), $\alpha$ from Eq. (\ref{alpha}), and $\gamma$ from Eq. (\ref{gamma}) yields:
 
\begin{equation}
\label{Trush}
\alpha +2\beta +\gamma=2+(d+2-\eta-2d_f)\nu.
\end{equation}
Thus, relations (\ref{eta}) and (\ref{dl2}) impose $d+2-\eta-2d_f=0$,  validating (\ref{rush}) or vice versa. This indicates that the additional relations derived from differential fractal geometry provide a robust framework, capable of yielding consistent exponent values even when certain scaling relations fail.

\section{concluding remarks and future perspectives}

 {\bf Concluding remarks -} %
In this work, we employ exact differential fractional geometry to generalize the results obtained for lattices of integer dimension to lattices of non-integer dimension. We recover exactly the same law,  Eq. (\ref{eta}), which establishes that the Fisher exponent $\eta$ is the difference between $d$ and the Riesz fractal dimension $d_R$,  originating from the cluster structure formed at critical temperature $T_c$ ~\cite{Lima24}. Using Table \ref{Table1} and Fig. \ref{figfractal}, we compute  $d_R$,  $d_f$ and $\eta$ to verify our claims.  Additionally, we focus on the anomalous region $1 \leq d \leq 2$, see Table \ref{Table2}, to test the robustness of our method and scaling relations, even in cases where $\nu$ diverges. 

Contrary to the common assertion that mean-field theory describes critical behavior only for $d\geq d_c$, we argue that the correlation function when reformulated in terms of the Riesz fractal dimension $d_R$, can describe phase transitions for any dimension $d$. Importantly, as $d-d_R$ decreases with $d$, the upper critical dimension corresponds to $d=d_c=d_R$, where our approach agrees with traditional Euclidean mean-field theory. Furthermore, Eqs. (\ref{G2}), (\ref{dl}), (\ref{eta}), and (\ref{dl2}) remain valid even for non integer dimensions. 

It is worth noting that Eq. (\ref{eta}) was obtained for any continuous order parameter $\psi(r)$ in a second-order phase transition, making its relevance to Ising models significant but not exclusive. So far, we have discussed the phase transition at equilibrium within a perfect lattice of dimension $d$, which was our main objective for this work.

Naturally, these ideas can be extended to explore various scenarios, ranging from equilibrium to a wide array of non-equilibrium processes. 

{\bf Future perspectives beyond equilibrium -} There are two situations where it may be worth continuing this work
\begin{enumerate}
    \item Growth phenomena
    \item Non--equilibrium phase transition
\end{enumerate}
{\bf Growth phenomena-} In growth processes, two quantities play an important role in growth, the average height $ \langle h(t) \rangle $ and the standard deviation of the height, also referred to as roughness or surface width, defined as $w(L,t)= \left[ \langle h^2(t) \rangle - \langle h(t) \rangle^2\right]^{1/2}$. Here, the average is taken over space. The roughness is a very important physical quantity, as it controls many physical phenomena it~\cite{Edwards82,Kardar86,Barabasi95,Gomes19,Reis05,Rodrigues15,Alves16,Carrasco18,Krug92,Krug97,Derrida98,Meakin86,Daryaei20,Hansen00,Rodrigues24}. For many growth processes, the roughness $w(L,t)$ increases {over} time until it reaches a saturated value $w_s$, such that $w(t \rightarrow \infty)=w_s \propto L^\chi$, where $\chi$ is the roughness exponent.   Saturation occurs not under equilibrium conditions, but in a steady state in which $w$ remains constant.

The surface roughness exponent $\chi$ can be related to the interface fractal dimension $d_f$ through the expression~\cite{feder2013fractals,Barabasi95}
\begin{equation}
\label{boxes}
\chi=
\begin{cases}
2-d_f , &\text{ if~~ } d=1,\ 2\;, \\
d-d_f, &\text{ if~~ } d \geq 2\;.\\
\end{cases}
\end{equation}
It is noteworthy that  Eq. (\ref{eta}), which describes equilibrium phase transitions, and Eq. (\ref{boxes}), which applies to out-of-equilibrium growth for $d \geq 2$, are very similar. Both equations consistently relate critical exponents to the deviation between the lattice dimensions $d$ and a system-specific fractal dimension. This suggests a possible general relation applicable to other types of transitions, provided a well--defined fractal dimension exists in each case. This idea is intuitively corroborated by the fact that the critical exponents are typically fractional numbers and functions of the dimension.


{\bf Out of equilibrium phase transition -} Considering that out-of-equilibrium growth phenomena exhibit analogous relationships, we infer that a general underlying relationship may exist, provided each case's fractal dimension is well-defined. Therefore, we propose that this hypothesis be rigorously tested across a various systems, such as those with a finite geometry\cite{GomesFilho20,Kalosakas22,salman24}, dynamic phase transitions~\cite{Ziff86,Fernandes18,Santos24} and synchronized phase transitions~\cite{Kuramoto84,Pinto16,Pinto17,Gutierrez24,GUTIERREZ25}. In the latter, once the synchronized state is reached, the order parameter becomes constant, exhibiting a behavior similar to magnetization\cite{Kuramoto84}.\\


{\bf Acknowledgments -} This work was supported by 
the Conselho Nacional de Desenvolvimento Cient\'{i}fico e Tecnol\'{o}gico (CNPq), Grant No.  CNPq-01300.008811/2022-51  and Funda\c{c}\~ao de Apoio a Pesquisa do Distrito Federal (FAPDF), Grant No.\ 00193-00001817/2023-43. . 

\vspace{1.0cm}

\section*{References} 
\vspace{0.5cm}


\providecommand{\newblock}{}
\expandafter\ifx\csname url\endcsname\relax
  \def\url#1{{\tt #1}}\fi
\expandafter\ifx\csname urlprefix\endcsname\relax\def\urlprefix{URL }\fi
\providecommand{\eprint}[2][]{\url{#2}}


\end{document}